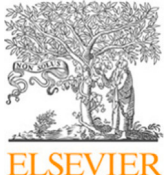

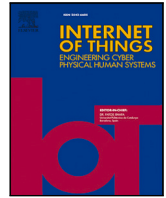

Research article

# Development and evaluation of Artificial Intelligence techniques for IoT data quality assessment and curation

Laura Martín *, Luis Sánchez, Jorge Lanza, Pablo Sotres

*Network Planning and Mobile Communications Laboratory. Universidad de Cantabria, Plaza de la Ciencia s/n, Santander, 39005, Spain*

## ARTICLE INFO

## ABSTRACT

Nowadays, data is becoming the new fuel for economic wealth and creation of novel and profitable business models. Multitude of technologies are contributing to an abundance of information sources which are already the baseline for multi-millionaire services and applications. Internet of Things (IoT), is probably the most representative one. However, for an economy of data to actually flourish there are still several critical challenges that have to be overcome. Among them, data quality can become an issue when data come from heterogeneous sources or have different formats, standards and scale. Improving data quality is of utmost importance for any domain since data are the basis for any decision-making system and decisions will not be accurate if they are based on inadequate low-quality data. In this paper we are presenting a solution for assessing several quality dimensions of IoT data streams as they are generated. Additionally, the solution described in the paper actually improves the quality of data streams by curating them through the application of Artificial Intelligence techniques. The approach followed in our work has been to append data quality information as metadata linked to each individual piece of curated data. We have leveraged linked-data principles and integrated the developed AI-based IoT data curation mechanisms within a Data Enrichment Toolchain (DET) that employs the NGSI-LD standard to harmonize and enrich heterogeneous data sources. Furthermore, we have evaluated our design under experimental research conditions, achieving a robust compromise between functionality and overhead. Besides, it demonstrates a stable and scalable performance.

## 1. Introduction

Proliferation of data sources associated to Internet of Things (IoT) deployment as well as those bound to Open Data Portals (e.g., European Data Portal, Municipalities Open Data Portals, etc.) and Social Media platforms is creating an abundance of information that is called to bring benefits for both the private and public sectors, through the development of added-value services, increasing administrations' transparency and availability or fostering efficiency of public services. In February 2020, the European Commission announced the European Strategy for Data [1], aiming at creating a single market for data to be shared and exchanged across sectors efficiently and securely within the EU. Behind this endeavor stands the Commission's goal to get ahead with the European data economy in a way that fits European values of self-determination, privacy, transparency, security and fair competition. This is especially important as the European data economy continues to grow rapidly — from 443 billion euros (3.6% of GDP) in 2021 to an estimated 787 billion euros (5.3% of GDP) by 2030 [2]. However, data itself is worthless (as gas would be if the combustion engine did not exist), data is only valuable as it supports intelligent decision-making processes and novel services.

* Corresponding author.
  *E-mail address:* lmartin@tlmat.unican.es (L. Martín).

Among the technologies that will play a key role in the forthcoming Economy of Data scenario, the Internet of Things (IoT) is recognized as a game-changer technology that expands its applicability to a huge variety of domains [3–6]. Precisely, the key asset is the data that the myriad of sensors embedded in the environment are constantly generating, which smartly processed can be transformed in added-value knowledge.

However, the larger the deployed IoT infrastructure, the more probable system and network failures are [7]. Large errors interrupting the system operation are handled by network management systems [8], but the widespread use of low-cost devices, which is critical to have wider scope and adoption of the IoT, raises several concerns especially pertaining to their accuracy, reliability, in-field applicability and performance [9]. For example, in the field of air-quality monitoring, there are hundreds of Low-Cost Sensors (LCS) commercially available on the market. However, although some reviews of the suitability of LCS for ambient air quality have been published [10], quantitative data for comparing and evaluating the agreement between LCS and reference data are mostly missing from the existing reviews. In general, they are less sensitive and less precise as their response is largely influenced by external factors and their behavior degrades with time [11]. The errors that they introduce can lead to poor Data Quality (DQ) that, in turn, lead to incorrect decision outcomes [12].

Although the IoT concept has been around for several years already [13,14], it still has many research and innovation open challenges, and even when there have been some previous works that have proposed methodologies and tools for DQ management or even ISO standards [15], the importance of IoT DQ has been somehow overlooked. Eventually, these works have concentrated on the Big Data aspect of the IoT, investigating, classifying and discussing DQ management for large datasets [16]. However, IoT has a fundamental dynamic nature and an important part of its potential comes from its ability to produce real-time data organized in so-called data streams (i.e. Time series of successive observations taken by IoT devices). Thus, it is of utmost importance that DQ is assessed and improved as the data is being generated.

The first challenge to overcome is to establish a clear focus and definition for DQ. In this respect, several definitions of DQ can be found in the literature [17]. Among them, there can be identified two different perspectives: one subjective, that focuses on measuring DQ from a practical perspective (i.e., evaluating it from a user perspective, emphasizing user satisfaction); the second one, objective, aims at evaluating DQ as a multidimensional concept where different dimensions of DQ, such as accuracy, timeliness, completeness, or consistency, can be elicited. Not all types of data show every of the identified dimensions, neither some of these dimensions are relevant depending on the application domain of the data. Thus, it is necessary to establish the DQ dimensions that should be evaluated. Moreover, it is critical to establish adequate mechanisms to perform this evaluation so that DQ assessment is correctly done.

However, assessing DQ is just a first step towards improving the usability of the data, which is the final aim. In this sense, the next challenge is to take action for actually increasing DQ. The first alternative that should be considered is to focus on the aforementioned subjective perspective of DQ and act on the usability of the data by linking the results of DQ assessment to the actual data. Thus, increasing the quality of the data on its informative axis and, overall, enriching the semantics of the data. This way, users (i.e. applications and services) consuming this data will not only have access to the data values themselves but also to the characteristics, along the evaluated DQ dimensions, of these values in the form of linked metadata. The second alternative is to compensate low marks in the DQ assessment process by implementing curation mechanisms that synthetically transform low quality data streams into higher quality ones. Thus, increasing the quality of data on its intrinsic axis.

In this paper we are presenting the work carried out in order to address these two alternatives. On the one hand, we are presenting the definition and the evaluation mechanisms to assess the dimensions that are relevant for IoT data streams DQ. In this sense, it is important to highlight that, even if we are focusing on data streams coming from IoT devices, the solution described in this paper could be used for other data streams sources (i.e. non-IoT data streams). However, as it has been already mentioned IoT has an impact on a large variety of application domains as a key source for such kinds of data streams and this is the reason for putting it as the focus. On the other hand, the paper presents the solutions that have been implemented in order to improve DQ of IoT data streams. The approach followed in our work has been to append DQ information as metadata linked to each individual piece of curated data. We have leveraged linked-data principles and integrated the developed AI-based IoT data curation mechanisms within a Data Enrichment Toolchain (DET) that employs the NGSI-LD standard [18] to harmonize and enrich heterogeneous data sources.

Thus, the key contributions that we present in this paper are: (i) identify and define the most relevant DQ dimensions in IoT data streams and the mechanisms to assess each of them; (ii) specify and implement data curation solutions that employ AI algorithms to enrich IoT data streams by increasing their quality features; and (iii) integrate these two kinds of mechanisms (i.e. assessment and enrichment of IoT data streams DQ) into an operational Data Enrichment Toolchain (DET) leveraging linked data principles and NGSI-LD standard to provide semantically-enabled enriched data.

The remaining of the article is organized as follows. In Section 2 we will be reviewing the related work that has been published in the field of IoT data streams DQ assessment and enrichment. The definitions of the DQ dimensions that has been assessed in our work and the description of the AI-based mechanisms that we have specified to carry out the assessment and curation of IoT data are presented in Section 3. In Section 4 we briefly introduce the DET and how we have integrated the DQ assessment and curation techniques within the whole toolchain in order to produce semantically-enriched IoT data streams. Section 5 presents the results of the evaluation analysis that has been performed using various techniques for IoT DQ assessment and improvement. Finally, in Section 6 we conclude the paper providing some closing remarks and highlighting the most important contributions of the work that we describe in this article.

## 2. Related work

The quality of the information in the context of research and development environments where data is the main element, is critical. Therefore, multiple definitions of the DQ concept can be found in the literature, being the most widespread the one proposed by [19], coining the "fitness for use" philosophy. In the same publication, a review of the criteria to be met in order to achieve the definition of DQ is carried out and a division of the possible quality dimensions into four main categories is made (i.e. Intrinsic, Contextual, Accessibility and Representation DQ). There are also numerous classifications, sets and definitions of the possible dimensions of quality, found and described in [19–21]. Reviewing all these classifications, [17] defines a basic set of DQ dimensions, while not achieving a general agreement on the description and calculation methodology of these features. Moreover, there are multitude of research studies that investigate not only the quality of the data itself, but consider the quality of the sources that produce such kind of data. Within this approach, [12,22] are examples driven by this strategy, emphasizing the need of addressing the poor quality directly from the source. However, we have focused our research in the quality of the actual data, as we stand by the idea that the data itself is the essential entity and, thus, where all the endeavors should be targeted. Focusing on DQ in IoT environments and data streams, in [23–25] a selection of DQ dimensions focused on this type of scenario (i.e. completeness, data volume, timeliness, accuracy, consistency and confidence) is made.

In our work we have followed the natural path of choice of DQ dimensions provided by the three publications mentioned above. Prioritizing the dimensions that provide information about the data individually and as a whole, we have selected accuracy, precision, timeliness, completeness and usability. We contribute this last dimension to this work, as it has not been defined before in the literature, to the best of our knowledge, and promotes the interoperability of the information and its easy consumption according to the homogeneous and standardized representation of the data streams.

In addition to the theoretical definition of DQ dimensions, it is necessary to specify the methods to perform actual quality assessment on data streams in IoT environments. In [26], authors focus on the calculation of a set of DQ dimensions and the application of some DQ improvement techniques on streaming time series, once a large enough dataset has been obtained to carry out these actions.

In our work, we have proposed the modeling and calculation of the selected DQ dimensions and several DQ enrichment techniques to be applied over the observations collected within IoT deployments in a continuous manner as they are generated. This difference makes our work stand out by enabling real-time management and monitoring of the IoT platform.

Moreover, once the DQ dimensions have been calculated and the DQ enrichment techniques have been applied, in our work we embed this information obtained as DQ metadata into the data streams. Hence, the maximum potential of data curation is achieved by being able to access these DQ features straightforward for each data stream. This approach is not widely used in the literature, and it is more typical to evaluate the quality dimensions periodically and send alerts if certain limits are exceeded, as proposed in [27].

Last but not least, in the literature, they can be found different research approaches regarding the integration of the quality assessment module in data acquisition and consumption architectures. In [27], an architecture and methodology for the evaluation and monitoring of DQ externally to an IoT platform is proposed. On the other hand, [28] discusses the addition of a set of activities that can be carried out in parallel or integrated with the other ongoing activities in the pipeline.

In our case, we have proposed the integration of the IoT Data Curation Module in an IoT platform with a DET architecture to probe the feasibility and suitability of embedding DQ dimensions in existing IoT platforms to further increase the Usability dimension that we have coined in this paper.

## 3. Data quality dimensions assessment

IoT environments generate a massive amount of information due to the myriad of connected devices. The quality of this information is a major issue and one of the main concerns in the use of this new technology due to its great impact on the final products [25,29]. This concept of DQ is crucial for information mining and analysis processes. DQ defines the degree of compliance with the requirements enforced by data consumers and DQ dimensions refer to the criteria that needs to be met for optimal analysis results and information consumption not being compromised. Some examples of DQ dimensions are the accuracy, reliability or error recovery [29].

In a complementary approach and with the purpose of enhancing these characteristics or DQ dimensions, methods or strategies are proposed to enrich not only the informative quality of the dataset, but also the quality properties themselves. That is, increasing or enhancing the values of these DQ dimensions through information processing techniques [29].

In this sense, it is important to highlight that we are not considering that the objective is to provide an absolute DQ ranking, but to provide insights about DQ dimensions that will be useful for consumers to understand the data items that they will be getting. Thus, it is not a matter of coming up with a compound value stating whether a data item within a data stream is of high or low quality, but to include metadata that allows deciding if that data item has enough quality or not, and this is something that is always subjective (i.e. it depends on the consumer requirements). The system that we are proposing does not go into the subjective interpretation of DQ, it only comprises the objective calculation and homogeneous semantic representation of DQ dimensions, so that data stream selection criteria can be enriched.

### 3.1. Data quality dimensions definitions

Recalling the definition of DQ dimensions as the quality criteria to be met in order to achieve the optimal results, these dimensions can be divided into four main categories: Intrinsic DQ refers to the inherent properties of the data itself; Contextual DQ stresses the importance of considering the context of use of the data; and Representational DQ and Accessibility DQ indicate the underlying requirements for data to be represented in a consistent and uniform way, and to be both easily understood and accessed [19], respectively.

Within a data streaming environment, in [25] five main dimensions are defined for assessing the quality of the environment and the devices. These dimensions are accuracy, confidence, completeness, data volume and timeliness. Accuracy indicates the maximum systematic numerical error produced in a sensor measurement, whereas confidence represents the maximum statistical error. Completeness addresses the problem of missing values or measurements in the dataset, data volume refers to the amount of underlying raw data and, finally, timeliness assesses the temporal context of the data stream. The first two dimensions described, accuracy and confidence, capture the quality attributes of the data individually (belonging to the category of "Intrinsic DQ"), as opposed to the remaining three dimensions, which refer to the quality of the data as a whole (belonging to the category of "Contextual DQ").

Taking this into consideration, in this work we have focused on the key Intrinsic dimension, which is the accuracy of the data, on three other critical Contextual dimensions (i.e. completeness, timeliness and precision), as well as on a fifth dimension, usability, which we have coined ourselves, that somehow compiles the Representational and Accessibility aspects of DQ.

In the following sub-sections, we are providing the definitions that we have used for them and their corresponding calculation methods.

#### 3.1.1. Accuracy

Accuracy indicates how close the measurement value is to the ground truth. As it can be seen in (1), it is the difference between the observed value and the reference value, taking the units of the observation value and represented by the symbol $\pm$ next to it.

$$accuracy = |observedValue - refValue| \tag{1}$$

#### 3.1.2. Completeness

Completeness quantifies the number of missed measurements or observations in a given time window. The method followed is shown in (2), using three variables: rate, inter-arrival time value; n, number of missed measurements observed; and window, time window of observation. This completeness parameter can be represented either as a part per unit or as a percentage.

$$completeness = \frac{window - n \cdot rate}{window}, \quad completeness\,(\%) = completeness \cdot 100 \tag{2}$$

#### 3.1.3. Timeliness

Timeliness is a dimension with a myriad of definitions within the literature [20,24,25,29]. The interpretation given by [29] describes the three main aspects affecting this property: the rate of update from the real-world to the system, the rate of change in the real-world, and the time when the data item is employed. In our approach, timeliness is purely concerned by the second one, and we have decided to base its calculus on a weighted average between the previous and the newly calculated mean update time value. This way consumers will get a grasp about the age and/or the punctuality of the data items. Eq. (3) shows this formula, where the following parameters are involved: $\alpha$, correction factor belonging to the range [0,1]; rawTimeliness, raw value of the newly calculated update time; meanTimeliness$_{i-1}$, mean value of the update time of the previous iteration; and meanTimeliness$_i$, mean value of the update time calculated in the current iteration.

$$meanTimeliness_i = \alpha \cdot meanTimeliness_{i-1} + (1 - \alpha) \cdot rawTimeliness \tag{3}$$

#### 3.1.4. Precision

This property does not belong to the original set of dimensions for IoT streaming assessment compiled in the literature, although it may appear in the background [17,19,20,23]. Taking as a basis the ISO 5725 definition for precision [30], our objective is to evaluate the closeness between the values in the data stream.

Following this approach, precision corresponds to the standard deviation of a dataset. That is, it measures how close the values in the dataset are to each other. Eq. (4) shows the standard deviation formula used, where the variables involved are: $\mu$, mean of the values in the dataset; n, number of samples in the set; and x$_i$, the $i$th element of the dataset. The precision result takes the units of the values in the dataset and is given with the symbol $\pm$ next to them.

$$precision = \sqrt{\frac{\sum_{i=1}^{n}(x_i - \mu)^2}{n}} \tag{4}$$

### 3.1.5. Usability

Opposite to the preceding ones, the Usability dimension has not been, to the best of our knowledge, previously defined, as such, in the literature [12,17,19]. The most similar concept might be the "fitness for use" described in [19]. It shares the approach of defining a multi-faceted quality dimension and the fact of being consumer-centric. However, it still focuses on the objective intrinsic dimensions and we think that it does not pay enough attention to the subjective and contextual dimensions related to the actual procedure of data consumption. As it focuses on the subjective perspective of DQ, it has not been qualified as one of the quality criteria that must be fulfilled to make data valuable. However, it is critically important to address three main aspects and potential dimensions that we have merged under the Usability dimension.

Firstly, the consistency of representation, that is, data has to be uniformly presented. For this, the use of standard information and data models is of utmost importance. Following open recommendations or schemes that have been backed by large communities of field-domain or cross-domain experts and practitioners makes the data more interoperable and easier to consume.

Secondly, the provenance and lineage of the data. Understanding where each piece of data comes from and which has been the pre-processing that it has been result of, is critical for consumers to have the necessary confidence to inject this data into their applications or to apply, or not, themselves further processing to it.

Finally, the informativeness of the data. Representing data in a semantically rich manner so that pieces of data are not just values but whatever associated characteristic to that value, particularly, those related to its quality dimensions is linked to it, would very much enhance the way in which that data is consumed. Thus, in our Usability dimension we are promoting the use of widely accepted information and data models supporting the capacity of expressing data in a semantic and linked manner as the way to have data that is more interoperable and easier to consume, thus, data of higher quality.

As this is a dimension in the subjective perspective of DQ, there is no closed expression to evaluate it, but for data to qualify as high-quality in Usability terms, it has to use standard information models, provide information to trace the origin and processes that the data has passed and make use of semantics and linked-data principles to attach relevant metadata to each piece of data.

### 3.2. AI-based mechanisms for data curation

In [29], DQ enhancement techniques are proposed for data provided in an IoT environment: outlier detection, interpolation, data integration, data deduplication and data cleaning.

Anomaly detection is one of the most well-known information enrichment techniques and it involves detecting hidden patterns that differ from the rest of the data. Once these anomalous points have been detected, there are two alternatives: either to suppress them or to highlight them. The second approach is used to search for unusual events, such as a fire. After this anomaly detection process, the accuracy and reliability of the resulting dataset is increased [29]. This DQ enhancement technique relies on two detection methods: outlier detection and novelty detection. The former is based on the detection of anomalous points within the training dataset, and the latter on the detection of these anomalous points in a previously unobserved dataset [31]. The most widespread methods for anomaly detection and identification are based on Artificial Intelligence techniques. Depending on the detection method to be used (outlier or novelty), the type of learning of Machine Learning algorithms must be different.

The type of learning focused on outlier detection is the unsupervised learning. These types of algorithms are based on clustering the data without prior labeling. That is, the training dataset is not classified and labeled, so the models are left to their own devices to detect and identify the structures and patterns in the data [32]. Within this type of learning, two of the most popular algorithms are Local Outlier Factor (LOF) and Isolation Forest (iForest). The first one, LOF, is based on calculating the deviation of the local density of a point regarding its neighbors, considering as an anomaly those that have a considerably lower density value than the rest. This local feature is achieved by focusing the evaluation of the outlier factor on a restricted neighborhood [33]. The second algorithm mentioned, iForest, involves the isolation of points in the dataset through the building of tree structures. Thereby, the anomalous points are those found in the shortest branches. It is worth noting that this algorithm presents a high detection efficiency by using a small portion of the dataset for tree generation. Therefore, it has low memory usage, which is ideal for working with high volume datasets [34].

The novelty detection approach is mainly based on semi-supervised learning. This type of learning involves training datasets that are classified into a single label. In terms of outliers, one might say that this is a clean dataset, without anomalies. Therefore, the trained model can detect whether the input dataset belongs to the known label or deviates from it [35]. A typical application of this type of learning is found in the One-Class SVM algorithm, derived from the SVM classification algorithm. This evolution allows the classification of the dataset into a well-characterized label or target class, without containing samples from another class or so-called outliers. Thus, novelty detection is performed by mapping and comparing with this target label [36].

However, unsupervised and supervised learning algorithms can be adapted to perform novelty detection. The latter type of learning is based on generating a predictive model from a training dataset that is already labeled [35].

Addressing the missing values in a dataset is also one of the most relevant enrichment techniques in information processing. It is crucial to deal with this question as missing values can bias the results of the techniques applied to the dataset or even reduce the quality of the information [29]. Focusing on value imputation, the two most important techniques are interpolation and Artificial Intelligence methods [12,29,37,38]. Interpolation is based on fitting the progression of the known dataset to a function of a given degree. This allows the value of the missing data to be determined. Artificial Intelligence based techniques used to deal with missing values are based on estimating these absent values according to the available information of the existing values. The algorithms used for this purpose primarily consist of supervised algorithms, of which kNN is one of the most significant. This algorithm, kNN, entails the classification of the values in the dataset into clusters or categories given their k nearest neighbors. Following this approach to absent value imputation, the k nearest neighbors of the missing value are used to estimate its value based on an inter-instance distance metric [37].

Within an IoT environment, where information is distributed as a time series and must be processed in real time as it is generated, some of the DQ techniques or dimensions mentioned above cannot be considered. Techniques using Machine Learning algorithms would need to retrain their models for each new observation received from the devices, as the timestamp of these observations is crucially important regarding the significance of their value. For this reason, one can argue that this continuous re-training, for each new piece of data collected, is not scalable. Hence, the estimation of future values is proposed to synthetically increase the dataset in order to be able to perform these DQ enrichment techniques in this additional time range. One of the most common approaches to forecasting future values is grounded in stochastic methods. Within the information generated by an IoT environment, the basic structure of these data streams is a time-series, obtaining measurement values on a periodic time basis. This data structure may present characteristics such as seasonality, which can be of a daily, weekly, monthly or annual nature. Due to this, the most suitable model for analysis and forecasting is Seasonal ARIMA (SARIMA) [39].

## 4. Autonomous IoT data streams curation

As it was introduced in Section 3.1.5, besides the objective DQ dimensions that we have focused on (i.e. accuracy, precision, completeness and timeliness), for whom the specific mechanisms developed to assess and enhance them have been already described in Section 3.2, the Usability dimension imposes additional requirements to make these aforementioned mechanisms worth. In this sense, we argue that applying DQ assessment or data curation techniques to IoT data streams is not worth itself. It will only become valuable if the outcomes of these techniques can be embedded into the actual data stream. This way, consumers will have access to higher-quality information (resulting from the data curation techniques) whose DQ features (resulting from the DQ assessment techniques) are known.

This information is consumed by applications offering different services according to a variety of analyses and processes. In order to ensure that the results obtained and the services provided are useful, the information employed requires to be as high quality as possible, both intrinsic and informative. Thus, the consumer is empowered to decide on the data to be used by considering the metadata attached to that information.

The approach that we have followed to embed the outcomes from the developed data curation mechanisms and introduce associated metadata to the information provided by IoT environments is through the integration of these mechanisms within a complete semantically-enabled DET.

In the following sections we are presenting the overall architecture of the DET and the key aspects that we have had to fulfill in order to integrate the data curation mechanisms into it. Particularly, the one related with the information modeling imposed by the NGSI-LD standard, which was employed to satisfy the Usability dimension in terms of using widely accepted information and data models, as well as the one pertaining to the orchestration of the different techniques into an integrated IoT data curation module.

### 4.1. Data enrichment toolchain architecture

In our work the DET is defined as the composition of heterogeneous microservices which results in the progressive enhancement of the original information quality and value. Conceptually, the DET can be understood as a pipeline with different sets of components, each one targeting a specific step within a particular data source improvement cycle.

The objective of the DET is to support the needs of the applications. To satisfy this requirement, the following functions implemented as microservices have been identified as key enablers of the architecture:

- data discovery, i.e., the ability to discover and request the collection of sets and streams of data;
- data formatting, i.e., the transformation of raw data into well-formed and structured set of data accordingly to data models described in terms of NGSI-LD;
- data curation, i.e., the identification (and potential correction) of data that do not reflect the expected quality (outliers, errors in values and the like);
- data linkage, i.e., the ability to relate different dataset accordingly to well established definition of relationships;
- data enrichment, i.e., the ability to understand and frame the data structures according to situations and contexts and the definition of functions that exploit this contextualization.

As it is shown in Fig. 1, heterogeneous data sources (e.g. batch, continuous stream, CSV, JSON, etc.) are discovered and crawled at the first step of the DET. Next, proper NGSI-LD mapping and transformation is applied in order to harmonize the incoming data and enable that subsequent processing can leverage semantics and linked-data principles. Before the data is stored into the Context Management Broker (i.e. the centerpiece of the NGSI-LD API [18]), data curation is applied in order to maximize DQ (both intrinsic and informative) so that the baseline for applications and remaining linking and enrichment functional blocks is as high-quality as possible information.

The AI-based mechanisms that we have developed represent the logic behind the data curation step of the DET. However, for them to be integrable within the toolchain it is necessary to first define the data models so that DQ-related attributes can be linked to the raw data and, second, orchestrate the execution of each of the aforementioned mechanisms so that the DQ of the NGSI-LD entities resulting from the processing is maximized.

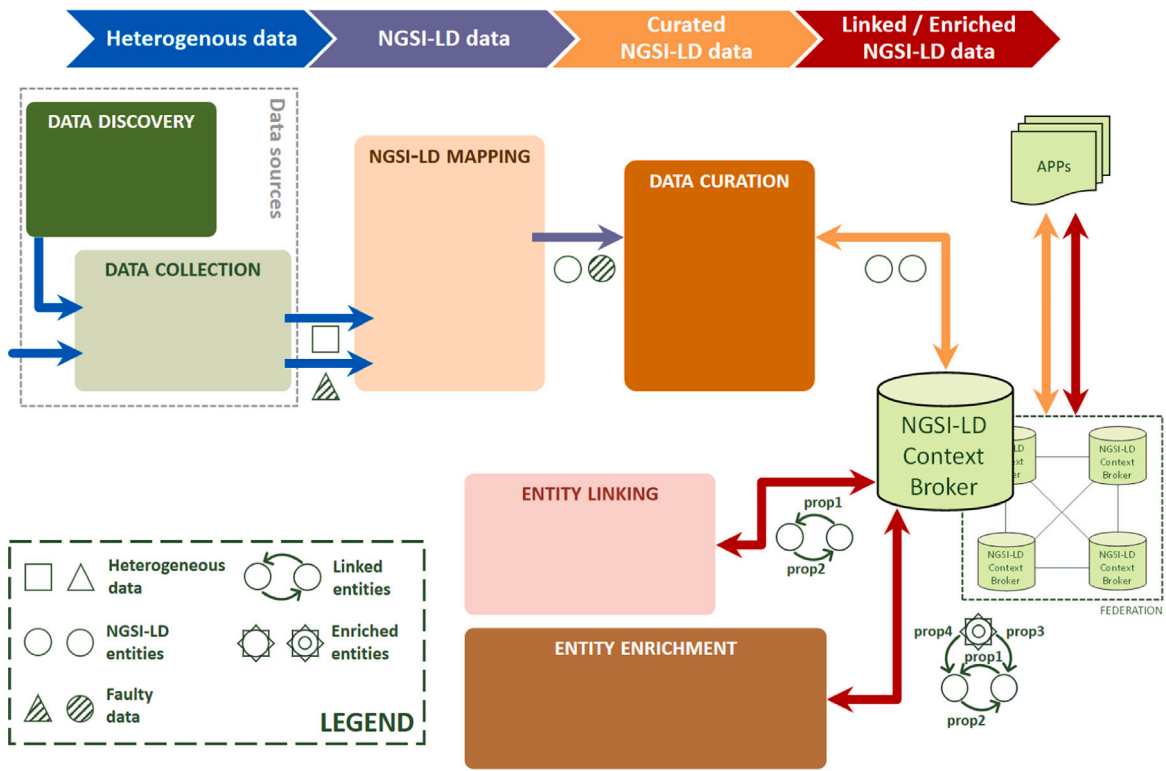

**Fig. 1.** DET high-level architecture.

## 4.2. Data quality information modeling

As it was concluded in Section 3.1.5 the outcomes from the application of DQ enrichment techniques and the analysis of DQ dimensions only become valuable if embedded into the data streams as metadata. Thereby, consumers are enabled to understand in a holistic manner the real information meaning not only from its raw value but also from its DQ features.

Therefore, we have decided that not only the measurements generated by the IoT devices, but also their DQ dimensions and properties, obtained with the application of DQ enhancement techniques, will be represented in the format defined by the NGSI-LD standard. The application of the NGSI-LD standard [18] for the representation of data streams and their DQ features allows the targets set in the Usability dimension to be achieved. It enables the uniform representation of information according to the standard, the traceability based on the principles of linked data and the provision of additional significance with metadata through property fields.

Since there was no specific data model available for representing the DQ assessment information at the Smart Data Models [40] catalogue, we have contributed to the initiative with a new data model to gather all the DQ dimensions addressed, as well as the properties acquired with the curation techniques for the traceability of their application. All the relevant files and documentation can be found in [41].

A schema of the contributed data model is shown in Fig. 2. Besides the DQ features, the basic fields of an NGSI-LD entity (*id* and *type*) are included, as well as a *source* in order to identify the ownership or source of the information and *dateCalculated*, pointing out when this DQ assessment was performed. The remaining properties correspond to the DQ features.

Beginning with those related to the objective DQ dimensions, there is one property for each of them: *accuracy*, *completeness*, *timeliness* and *precision*. In turn, all of these are described as objects with four internal fields: *type*, *value*, *observedAt* and *unitCode*. The *type* property is specific to the NGSI-LD information model, *value* represents the numerical value (e.g. float) of the dimension, *observedAt* allows keeping a log of the time at which this property has been analyzed and, lastly, *unitCode* indicates the unit code of the measurement.

As can be seen, there are two alternative ways of keeping temporal logs: with the *dateCalculated* property and with the *observedAt* subproperties. The former one allows the user or the data producer to record the DQ assessment entity as a whole, whereas the latter one belongs to each property, which is used to keep track of that property individually. Thus, including both logging methods enables the data model to be used widely, that is, by several use cases.

Furthermore, the *outlier* and *synthetic* properties relate to the applied DQ enhancement techniques employed over this piece of data, so that it is possible to trace back which pre-processing has been made over it. The first one, *outlier*, indicates within its

```
{
    "id": <string>,
    "type": <string>,
    "source": {
        "type": "Property",
        "value": <string> },
    "dateCalculated": {
        "type": "Property",
        "value": <datetime> },
    "accuracy": {
        "type": "Property",
        "value": <float>,
        "observedAt": <datetime>,
        "unitCode": <string> },
    "completeness": {
        "type": "Property",
        "value": <float>,
        "observedAt": <datetime>,
        "unitCode": <string> },
    "timeliness": {
        "type": "Property",
        "value": <float>,
        "observedAt": <datetime>,
        "unitCode": <string> },
    "precision": {
        "type": "Property",
        "value": <float>,
        "observedAt": <datetime>,
        "unitCode": <string> },
    "outlier": {
        "type": "Property",
        "value": {
            "isOutlier": {
                "type": "Property",
                "value": <boolean> },
            "methodology": {
                "type": "Relationship",
                "object": <string> } },
        "observedAt": <datetime> },
    "synthetic": {
        "type": "Property",
        "value": {
            "isSynthetic": {
                "type": "Property",
                "value": <boolean> },
            "methodology": {
                "type": "Relationship",
                "object": <string> } },
        "observedAt": <datetime> }
}
```

**Fig. 2.** Schema of the contributed Smart Data Model for DQ assessment.

internal fields whether or not the measurement corresponds to an anomalous value (*isOutlier: True/False*), whereas also including a *methodology* field to provide information on the AI method used to come to this conclusion. The latter, *synthetic*, determines through the *isSynthetic* Boolean property whether it was a missing observation in the time series which has been created synthetically using AI. A *methodology* field is included, as for the *outlier* property, in order to add traceable parameters of the application of these techniques.

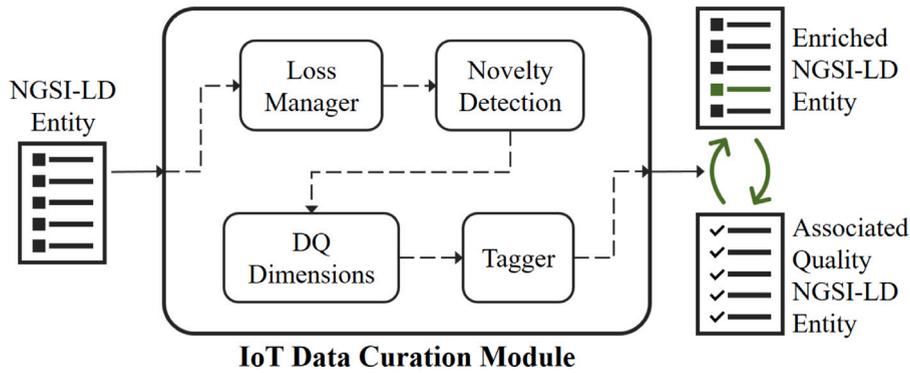

**Fig. 3.** Internal architecture of the IoT data curation module.

In order to link the actual data value with its DQ features, the modeling that we are proposing would link a new field to the data models used for the representation of the streams. This way, the value entity (data stream) and the associated quality entity (DQ features) will be explicitly bound.

### 4.3. IoT data curation module

The component responsible for information curation within the Data Enrichment Toolchain (DET) architecture is called the IoT Data Curation Module. Its main objective is to apply the DQ enhancement techniques and to obtain the DQ dimensions to be able to store the information with the highest possible quality into the Context Management Broker. Its internal architecture is depicted in Fig. 3, along with the required input element and the resulting output elements. It can be seen that these elements are represented in the format defined by the NGSI-LD standard, as discussed in the previous section. The data streams are mapped into the most appropriate Smart Data Models available and the DQ features into the properties of the new data model proposed for DQ assessment. It is worth noting that a new property must be added to the value entity (data stream) that relates this entity with its DQ features.

The internal process followed by the curation module is based on four key steps.

First, the Loss Manager sub-module is in charge of identifying whether a loss has occurred between the reception of the current measurement and the last one logged. When no loss has been detected, this module behaves as a NO-OP (no operation), continuing with the next step. However, if it is found that a loss has occurred, the internal logic of this component will generate a synthetic measurement (NGSI-LD entity) on which the remaining techniques will be performed. In this way, the Context Management Broker will keep a complete record of all the readings produced by a periodic IoT device, knowing whether it is a real or synthetic measurement (labeled in the contributed data model and discussed in the previous section).

Thereafter, novelty detection is performed. Using the dataset forecasted for the short-term future, it is determined whether the value of the observation at its timestamp (NGSI-LD entity) corresponds to an anomaly.

The last quality assessment sub-module is dedicated to the acquisition of the DQ dimensions. Using the methods discussed aforementioned and the necessary values of the entities stored in the Context Management Broker, the DQ features of the NGSI-LD entity being assessed are characterized.

Finally, once all the proposed DQ features have been obtained, they must be tagged in the new NGSI-LD entity following the DataQualityAssessment data model presented in Section 4.2. To this end, the properties are filled in with the values collected and a new field is included in the NGSI-LD value entity (the one being assessed). This allows both entities to be related and to benefit from the quality assessment and tagging process.

## 5. Data quality enhancement techniques evaluation

In this section we are describing the evaluation that has been carried out to characterize the behavior and performance of the proposed DQ enhancement techniques. In Section 5.1 we assess the overhead that evaluating and introducing DQ-related metadata imposes, how it evolves in time, and we discuss its worthiness. Moreover, in Section 5.2, we are presenting the evaluation of the algorithms and methods proposed to perform DQ enhancement techniques (i.e. outlier and novelty detection and data item estimation).

### 5.1. Specifics of calculation and performance evaluation of the DQ dimensions

Section 3.1 discussed the DQ dimensions definitions that we have followed and proposed in our approach. In order to characterize their behavior an evaluate their performance, we have set up a testbed environment in which these DQ dimensions' calculations have been carried out.

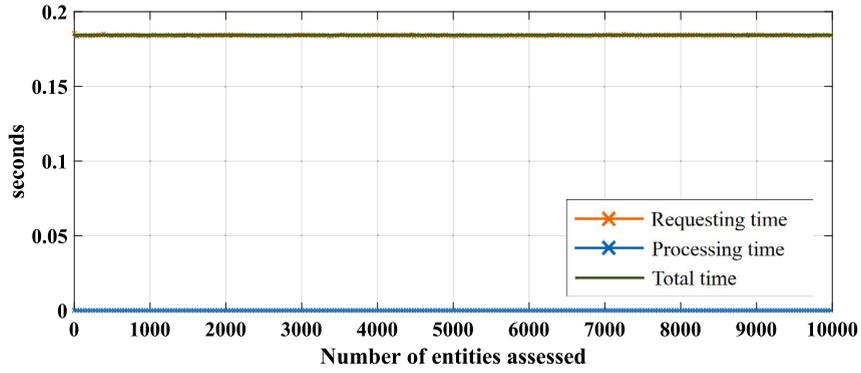

**Fig. 4.** Accuracy performance. It shows the median values of the 60 Monte Carlo simulations.

We have simulated the existence of 100 different sensors which generated observations every two minutes. Also, the simulation has been running until each sensor had generated 100 values. With this, we had a data stream including 10 000 data items as input to our DQ assessment toolchain. For this purpose, we have used the measurements in a dataset of historical temperature records from SmartSantander [42] sensors (both mobile and fixed) from 01 January 2021 to 13 June 2022. Furthermore, we employed the Monte Carlo method [43] for the evaluation of the performance of each of the DQ dimensions analyzed. Following this experimentation method, 60 rounds of this same environment were executed for every experiment in order to allow us deriving generally valid conclusions.

All these tests have been carried out on an Ubuntu 20.04.5 LTS machine (2 CPU cores, 2.40 GHz clock, 16 GB RAM) in which a Scorpio NGSI-LD Broker has been used as the Context Management Broker. After each of the 60 rounds, the Scorpio Context Broker was re-started and put into clean slate.

The formula used for calculating the overhead, in terms of size, introduced by the inclusion of the metadata generated by the quality assessment procedure is presented in (5). Therein, *enriched* refers to the entity size including insights about DQ dimensions, and *raw* refers to the original size of the entity (i.e. only its value without additional insights).

$$overhead(\%) = \frac{enriched - raw}{raw} \cdot 100 \tag{5}$$

As previously mentioned, all relevant files and documentation on the DataQualityAssessment Smart Data Model are available at [41]. Together with the specification files, it is possible to find examples of its use and application.

Additionally, the calculation of each dimension also introduces some computation time overhead. This overhead can be split in the necessary time to get all the information required to perform the calculation of the DQ dimension for each data item (so called, *Requesting time*), and the necessary time to actually execute that calculation (so called, *Processing time*).

In the following sections, we are reviewing the overhead introduced both in terms of size and in terms of computation time for every dimension individually.

### 5.1.1. Accuracy

Recalling the definition of accuracy, its value is obtained with respect to a reference value. Thus, a request to an external trusted source has to be made, and, consequently, the accuracy calculation workflow consists on two phases: requesting the reference value and processing the calculus.

For each measurement received, the Curator module (specifically the DQ Dimensions sub-module) makes a request to an external trusted source considered as ground truth (if any) querying the most recently observed value. For the purpose of this research experiment, the external trusted source corresponds to the Spanish State Meteorological Agency (AEMET) [44]. Once the sub-module receives this information, it performs the corresponding calculation (cf. Eq. (1)).

Fig. 4 presents the results of the computation time evaluation. The three aforementioned times (i.e. *Requesting time* in orange, *Processing time* in blue and, finally, *Total time* in green) are shown. It is worth noting that the median values of the 60 Monte Carlo repetitions are represented so that the occurrence of occasional abnormal values does not distort the figure. However, in Table 1 we present the mean and standard deviation values of these three temporal variables. As it can be seen, the negligible value of the $\sigma$ demonstrates the correctness of representing the median values, as well as the flatness of the Curator module's time response.

The accuracy *Total computation time* shows a completely flat response. *Processing time* can be neglected as it is almost null. Thus, the performance is totally dependant on the response time of the external trusted source. Though, it is important to highlight that this performance is stable with the time (i.e. it is not altered with the number of data items in the data stream), so system's scalability is granted.

Lastly, the calculation of the accuracy dimension results in an extra quality metric to be added to the raw data. In this case, the raw data (Temperature entity) is 1205 bytes in size, and the accuracy property consists of 134 bytes extra. Using the expression shown in (5), the metadata related to the accuracy dimension insights entails an overhead of 11.1%.

**Table 1**
Metrics for the evaluation of the accuracy performance.

|  | $\overline{X}$ (s) | $\sigma$ (s) |
|---|---|---|
| Requesting time | 0.1851 | 0.0056 |
| Processing time | 4.0877e−06 | 2.7319e−07 |
| Total time | 0.1851 | 0.0056 |

**Table 2**
Metrics for the evaluation of the completeness performance.

|  | $\overline{X}$ (s) | $\sigma$ (s) |
|---|---|---|
| Requesting time | 0.0407 | 8.4879e−04 |
| Processing time | 6.1895e−05 | 9.5658e−07 |
| Total time | 0.0407 | 8.4887e−04 |

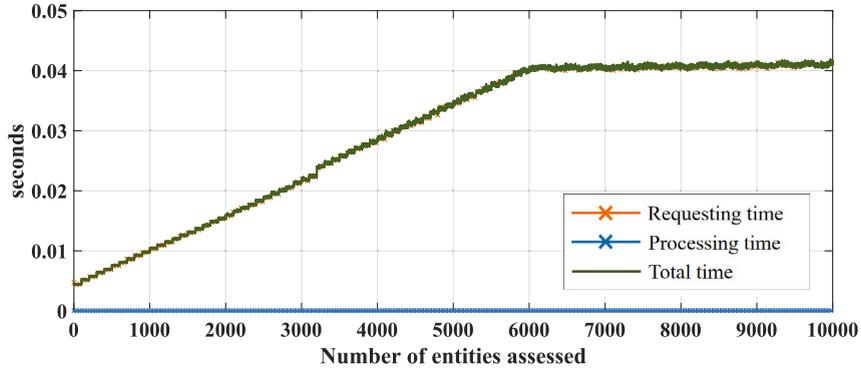

**Fig. 5.** Completeness performance. It shows the median values of the 60 Monte Carlo simulations.

### 5.1.2. Completeness

In order to perform the completeness dimension calculation, it is necessary for the Context Broker to support the storage of temporal values. That is, support having access to the historic data items generated by the same sensor, which are stored as temporal values associated to each data measurement entity. Analogous to the accuracy dimension calculus, the process can be split in two phases: first, getting the necessary temporal values from the Context Broker and, second, performing the actual calculation.

Thus, the first step is to request the temporal values stored in a pre-defined time window for each observation received. This pre-defined time window enforces the temporal correlation for the completeness dimension assessment, meaning that this is a dimension that should be looking to a limited history and not to the complete history of the data stream. The sub-module has to query these historic values for both the entity type assessed (e.g. Temperature) and the DataQualityAssessment entity linked to it. After obtaining these data, the sub-module is capable of evaluating the expression in (2). It is important to note that $n$ would be the number of values tagged as synthetic (*isSynthetic = True*) and rate would be the timeliness value of the current observation.

The results of the evaluation experiments are shown in Fig. 5. As in the previous case, both the median values of the *Requesting time* and the *Processing time* are presented together with the median *Total computation time*.

The completeness computation presents a characteristic response. Recalling the simulation environment where we had 100 different sensors generating values every two minutes, and having set the aforementioned pre-defined time window for the request to 120 min (i.e. 60 temporal entities), it can be seen that there is a turning point when the Context Management Broker is able to return these 60 temporal values for each of the 100 sensors. That is, the slope represents the increase in the response time as more entities are posted into the Scorpio until it reaches the 60 temporal entities. Hereafter, the temporal response becomes flat reaching an stationary behavior, although new values are being posted. This performance shape is due to the filtered query done to the Context Broker, which is requesting just the last 60 temporal values (120 min of time window). On the other hand, regarding the *Processing time*, it can be concluded that its impact is negligible.

Though, Table 2 presents the mean and standard deviation of all these times for the stationary phase, as this should be the overhead introduced on the long-term execution of the proposed system. As it happens in the previous case, it is remarkable the small standard deviation presented by the system performance, highlighting the scalability of the proposed module.

Finally, we have calculated the overhead imposed by the addition of the completeness property to the raw entity. Recalling the size of the initial entity, which was of 1205 bytes, knowing that this new property implies 137 bytes and using the equation in (5), we obtain an overhead of 11.4%.

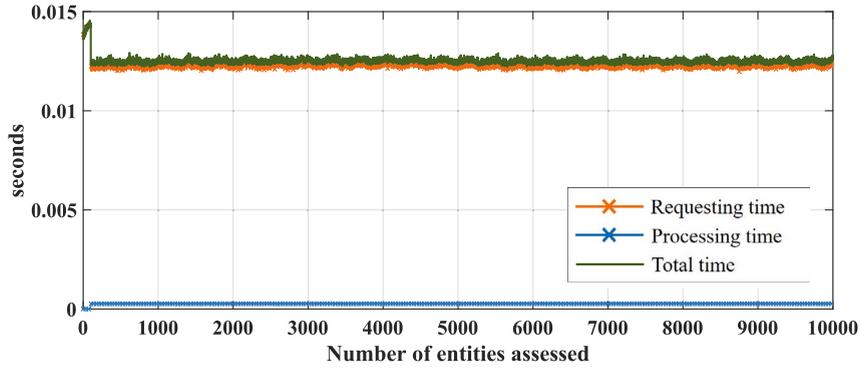

**Fig. 6.** Timeliness performance. It shows the median values of the 60 Monte Carlo simulations.

**Table 3**
Metrics for the evaluation of the timeliness performance.

|  | $\overline{X}$ (s) | $\sigma$ (s) |
|---|---|---|
| Requesting time | 0.01223 | 4.7811e−04 |
| Processing time | 2.6681e−04 | 3.7468e−06 |
| Total time | 0.01250 | 4.7807e−04 |

### 5.1.3. Timeliness

Similarly to the previous cases, the procedure that we have followed for calculating the timeliness dimension (considering the interpretation that we have described in Section 3.1) can be split in a two-step procedure (i.e. requesting the information and calculating the DQ dimension value).

In the requesting phase, the DQ Dimensions sub-module queries the Context Broker for the last value recorded for the DataQualityAssessment linked entity every time it receives a new measurement. Once the module collects this piece of information, it can execute the calculus as in (3).

Fig. 6 depicts the performance of the timeliness computation. The two phases are assessed in terms of the delay that they introduce, as well as the sum of these two (median *Requesting time* in orange, median *Processing time* in blue and median *Total time* in green, respectively). The behavior during the experiments was always the same except for the very first 100 data items in the data stream. For this initial part of the experiment a small relative performance degradation can be seen, but for the long-term run of the experiment, the behavior is completely flat. This transitory part at the beginning of the experiments is due to the behavior of the Context Broker when it is requested to provide a fixed number of entities that do not yet exist (recall that the Scorpio is re-started and put into clean slate for each round of the experiment). When this happens, there is this small degradation that disappears the moment there are enough entities to completely satisfy the query that the Curator module makes during the first of the aforementioned two-stepped timeliness calculation procedure.

Table 3 presents the mean and the standard deviation of these three computation delays. The practically null contribution of *Processing time* to the sum of the total is noteworthy, as well as the insignificant $\sigma$ shown in all variables.

These results allow us to conclude that the scalability of our system regarding the calculation and processing of the timeliness dimension is fully guaranteed.

Before ending with the timeliness dimension, it is necessary to calculate the overhead entailing this property. As in the previous cases, the raw data comprises 1205 bytes, and the metadata associated to the timeliness dimension adds 140 bytes. Thus, as from (5), the overhead introduced is of the 11.6%.

### 5.1.4. Precision

The last dimension that was processed in the system that we are presenting in the paper is precision. Regarding the definition given earlier in Section 3.1, once again the calculus process consists of two parts. The first one concerning the requesting operations and the second one related to the calculus itself.

Thus, for each measurement received, the Curator module requests from the Context Broker the latest information about all the entities within a distance range (i.e. enforcing both temporal —latest value—, and spatial —geographically close sensors— correlations), for both the entity type assessed and the DataQualityAssessment type linked to them. After getting all this information, it can start assessing the precision of the received data item. The DQ Dimensions component selects the values tagged as inlier (*isOutlier = False*) in its corresponding/linked data quality entity. With all this, it computes (4) and obtains the precision value. It is worth noting that, in order to calculate the precision of a value within a data stream, this assessed value has to play the role of the $\mu$ parameter in (4).

Similarly to the previous dimensions, the median delay is presented in Fig. 7 and it is split into the two phases of the dimension's assessment procedure (*Requesting time* in orange, *Processing time* in blue and *Total time* in green).

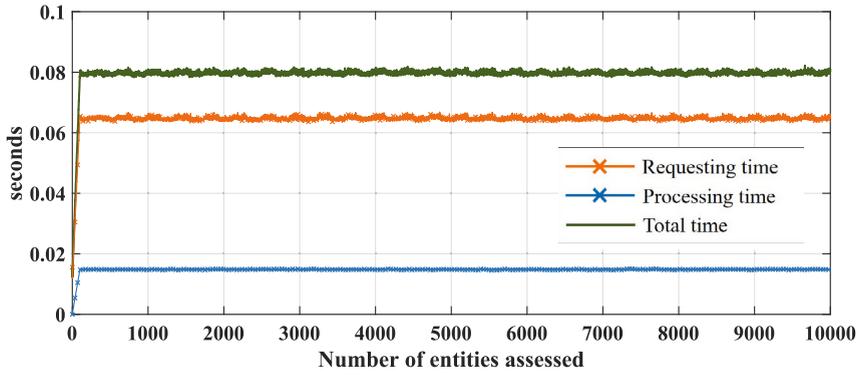

**Fig. 7.** Precision performance. It shows the median values of the 60 Monte Carlo simulations.

**Table 4**
Metrics for the evaluation of the precision performance.

|                 | $\overline{X}$ (s) | $\sigma$ (s)  |
|-----------------|--------|-----------|
| Requesting time | 0.0647 | 0.0012    |
| Processing time | 0.0148 | 1.0452e−04 |
| Total time      | 0.0798 | 0.0012    |

As it happened in the completeness computation, there is a transitory period that, in this case, lasts for the first 100 data items. In the first period the *Total computation time* grows linearly due to the nature of the precision dimension calculation. Since only 100 different entities (different sensors) are generated, this is where the breaking point occurs. From this point on, the response becomes flat. Therefore, also in this case, it can be concluded that the system has a high scalability since, despite the increase in the number of entities stored in the Context Broker, the *Requesting time* and the *Processing time* remain constant.

Table 4 gathers the mean and standard deviations values for all three variables.

Finally, the overhead introduced by the addition of the precision property as metadata has been calculated. Taking (5) as a basis, considering that the raw data is 1205 bytes and this new property adds 134 bytes, we obtain an overhead of 11.2%.

### 5.1.5. DQ metadata overhead discussion

As it has been described in the previous sections, the inclusion of the DQ-related metadata imposes a non-negligible overhead that might be considered as a drawback to the proposed solution. However, as from this analysis, it is clear that the increase in transmission and storage requirements, as well as the processing delay, due to the inclusion of the additional properties is not so big price to pay for enabling important insights about the available data.

Moreover, it is important to highlight that from the transmission viewpoint, DQ assessment is made at the DET level so the overhead is not affecting the uplink (i.e. the transmission from data source to DET) before the data is processed by the respective injection chain. Additionally, the NGSI-LD API allows for filtering the properties in which data consumers are interested. Thus, those consumers that are not interested in having insights about the quality features of the data that they are consuming, can simply filter them out and also the downlink (i.e. the transmission from DET to data consumer) will not include any overhead either.

From the storage viewpoint, the current design and implementation of the DET defines that metadata is computed within the injection chain so it is also stored at the Context Broker. Though, a possible optimization could be to include the metadata only at request time in case the consumers show their interest in getting such insights. Thus, not storing it but generating it when a consumer actually requests it and only for that consumer. However, in this case, the overhead would be the introduction of processing delay as the requested DQ dimensions would have to be computed on the fly. Moreover, this alternative would also require from altering the NGSI-LD API, thus breaking the standard specification, since the corresponding DQ assessment engines would have to be orchestrated upon a context data request arrives, or a notification to a previous context data subscription is triggered.

Having access to added-value functionalities (in our case, valuable insights about data quality) always come with some overhead. We think that the proposed solution is a sensible trade-off between functionality and overhead. Moreover, it has demonstrated stable performance and scalability.

### 5.2. Evaluation of AI-enabled data quality improvement techniques

Section 3.2 reviewed the DQ enhancement techniques used to increase the DQ dimensions and, therefore, the quality of the information. The aim of this section is to evaluate the algorithms and methods proposed to perform these techniques. Regarding the main objective of the Curator module, we recall its use to assess data streams. Therefore, in order to use AI techniques, it is required to train the engines prior to setting up the whole system. The high-quality dataset used to train the desired models has been prepared through the application of the Machine Learning algorithms in their version for static dataset. Thus, it is worth noting that there are two different but connected flows: first, a high-quality static dataset is generated, and then, this dataset is the one used as a training for the engines handling the data streams.

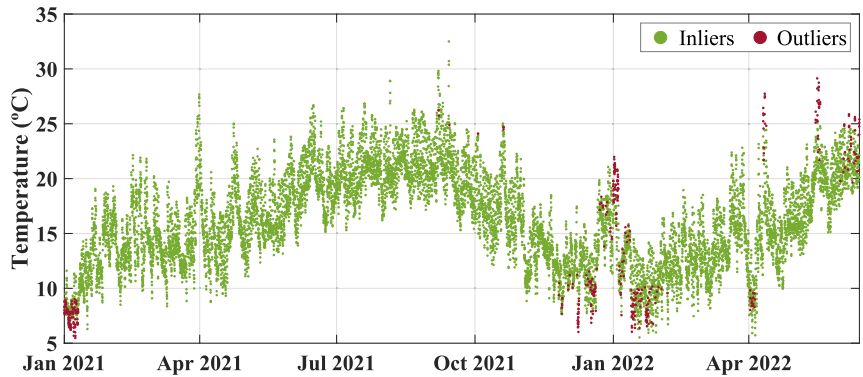

**Fig. 8.** Outlier detection on the Santander temperature dataset with the Isolation Forest algorithm (number of estimators = 200, contamination factor = 0.035).

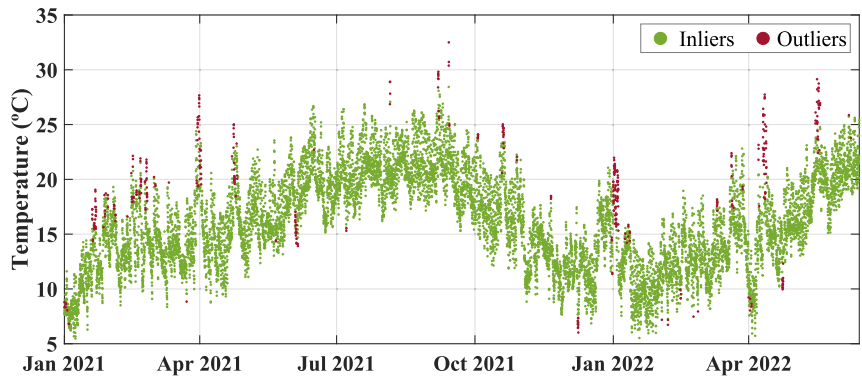

**Fig. 9.** Outlier detection on the Santander temperature dataset with the Local Outlier Factor algorithm (number of neighbors = 70, contamination factor = 0.035).

Starting with the static dataset, domain knowledge is applied for the sake of curating its values and format, and after that, we moved on to perform and evaluate the aforementioned enhancement techniques. For this purpose, we have used the same dataset of historical temperature records (from 01 January 2021 to 13 June 2022) from SmartSantander sensors that we employed in Section 5.1. As a Smart City project, SmartSantander has a huge number of sensors that report periodically every 1, 2 or 5 min, reaching a volume of around 2 GB of data, with approximately 16 million observations, during the period of time previously mentioned (01 January 2021 to 13 June 2022). In line with the premise of being a Smart City project, the quality of the devices deployed around the city of Santander is not high, leading to inaccuracies in the observations or even the recurrence of absurd values. Besides these problems, there is also the temporal degradation suffered by the devices, as they have been deployed since the beginning of the project in 2011. Consequently, it is unquestionable that the initial dataset must be processed prior to the application of the DQ enhancement techniques. According to this, three methods regarding the domain knowledge have been applied: elimination of nonsense, spatial correlation and temporal correlation.

The first method, elimination of nonsense, is based on setting logic limits to the possible values of the observations read by the sensors. In our case, as we are dealing with the temperature phenomenon in the city of Santander, we have consulted the AEMET [44] records in the time range that delimits the dataset.

The second method corresponds to the application of spatial correlation. Due to the fact that in the SmartSantander project some devices were placed in buses and taxis, reports have been registered outside Santander, for example in Bilbao or Madrid. Therefore, by focusing the study of the whole dataset within the geographical area of Santander, those observations taken outside these coordinates are deleted.

The last method of preprocessing the dataset is focused on temporal correlation. By grouping the observations in a periodic frequency (hourly in our case) and averaging their values, this aspect is achieved. Thus, the possibility of duplication of timestamps of observations from different sensors is eliminated and the volume of data is considerably reduced, without losing information.

Once the data have been preprocessed, the DQ enrichment techniques are applied to obtain a high-quality static dataset and to evaluate the methods and algorithms proposed in Section 3.2.

To this end, the first technique to be applied is the outlier detection, with a comparison of the results achieved after the implementation of the two aforementioned algorithms. Figs. 8 and 9 show the observations detected as outliers in red versus those considered as inliers in green with the following algorithms: Isolation Forest (iForest) and Local Outlier Factor (LOF), respectively. A difference can be observed between the behavior of the two alternatives. The anomalous points detected with the Isolation Forest algorithm correspond to global outliers within the dataset, whereas those detected with the Local Outlier Factor algorithm respond to local anomalies related to the group of neighbors chosen for evaluation. Although this difference has already been discussed theoretically in a previous section, through these results, we can verify it graphically. It is worth noting that the outcome obtained is affected by the choice of parameter values for each of the algorithms (mentioned in the captions of the figures).

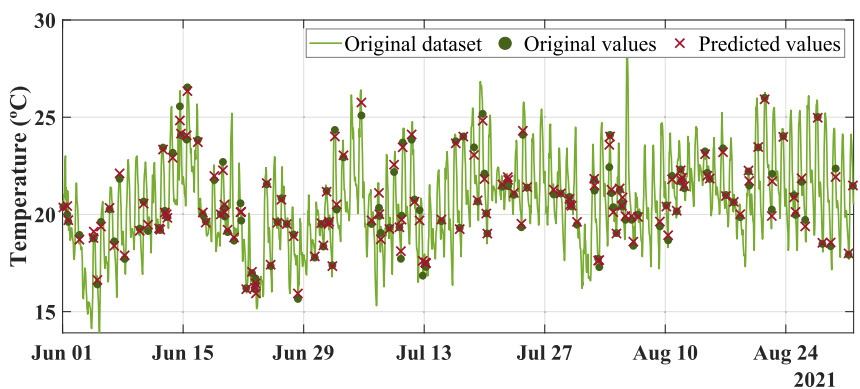

**Fig. 10.** Estimation of values with polynomial interpolation of degree 2 on a three-month dataset.

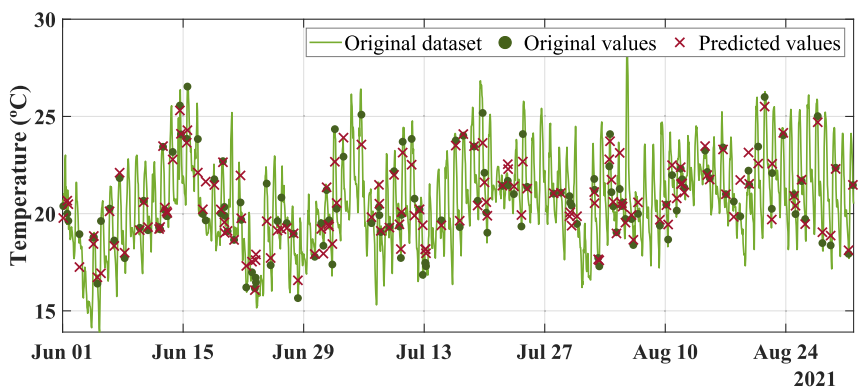

**Fig. 11.** Estimation of values with the kNN algorithm on a three-month dataset (number of neighbors k = 5).

**Table 5**
Metrics for the evaluation of value estimation methods.

| Technique | Mean Absolute Percentage Error (MAPE) | Mean Absolute Error (MAE) |
|---|---|---|
| Polynomial interpolation of degree 2 | 0.000592 | 0.0121 |
| kNN with 5 neighbors | 0.001740 | 0.03541 |

After extracting the set of outliers, we remove them from the overall dataset, resulting in missing values or gaps. Therefore, the next step consists of the imputation of values in these blanks into the dataset. For this purpose, the two formerly mentioned techniques are evaluated: interpolation, namely polynomial interpolation of degree 2, and the Machine Learning algorithm kNN, with the output shown in Figs. 10 and 11, respectively. Given the representation of the results, the dataset has been zoomed to a time range of three months in order to see the outcome more clearly. In order to evaluate both techniques, the dataset has been separated into two subsets: training set and validation set. The first one is composed by the dataset with gaps to be imputed and the second one consists of the values that were in those gaps, to be able to evaluate the accuracy of the estimations. The figures demonstrate the minimal difference in the performance of both techniques, representing the original values (validation set) with dark green dots and the estimated values with red crosses.

Table 5 shows the most common evaluation metrics of the estimation methods. It is noticeable that the polynomial interpolation of degree 2 performs better than the kNN algorithm, achieving precisions of 99.941% and 99.825% (1-MAPE), respectively. The difference is minimal, thus both techniques are good choices for value imputation.

After choosing the missing value imputation method, a clean and high-quality dataset is obtained. This is the outcome of this first flow, as mentioned at the beginning of this section. The next step, concerning the data streams flow, is to train the ML engines for novelty detection with this high-quality dataset. The algorithms chosen are the Isolation Forest, Local Outlier Factor and One-Class SVM towards the novelty detection. To this end, an unobserved dataset is generated, on which these AI procedures are going to be carried out. Fig. 12, Figs. 13 and 14 show the results obtained for each of the algorithms. Starting with the latter one, One-Class SVM, it can be observed that it does not work properly. This is due to the nature of the dataset, as this algorithm can only be applied on vector sets and not directly on time-series type of data. Instead, a conversion between the two types of data needs to be performed [45].

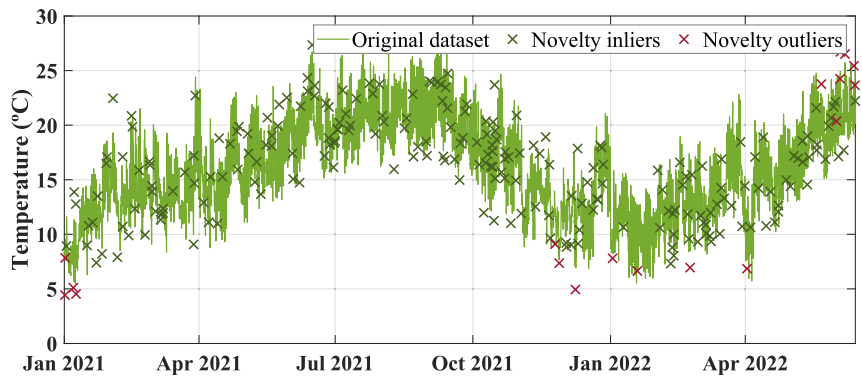

**Fig. 12.** Novelty detection on the Santander temperature dataset after anomaly cleaning and estimation of the remaining gaps with the Isolation Forest algorithm (number of estimators = 200, contamination factor = 0.035).

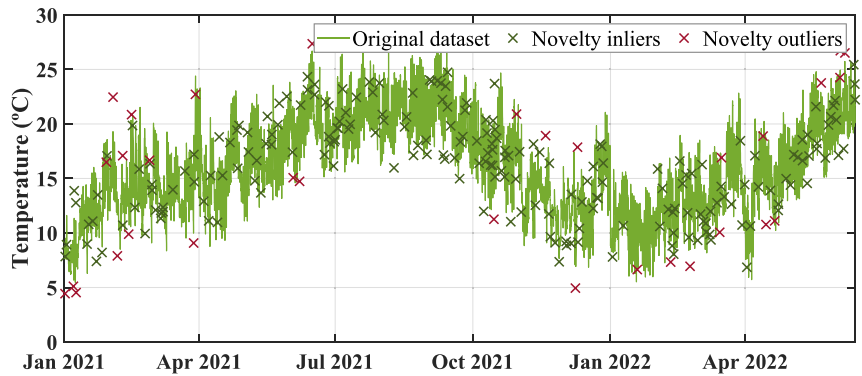

**Fig. 13.** Novelty detection on the Santander temperature dataset after anomaly cleaning and estimation of the remaining gaps with the Local Outlier Factor algorithm (number of neighbors = 70, contamination factor = 0.035).

As for the other two algorithms, differences can be observed in their results. Isolation Forest presents a more relaxed performance with the possibility of under-detecting anomalies, whereas Local Outlier Factor behaves in the opposite way with stricter results and the chance of over-detection. Therefore, the decision is left to the restrictions imposed by the application that makes use of this DQ enhancement technique.

The novelty detection just performed has been done on what is considered a static dataset in a known time range for the AI model. This has helped us in our efforts to evaluate the different performances of each algorithm. However, when it comes to an IoT environment where curation mechanisms need to be applied as observations are generated (as in our case), the clean and high-quality dataset obtained as outcome of the first flow does not have the temporal instants of these new observations.

Therefore, in order to perform real-time novelty detection for each of the data streams, this base dataset has to be synthetically extended with the intention of having that extra time range. The last DQ enrichment technique to be evaluated is this synthetic extension of the base dataset. Given the nature of the dataset (time series with seasonality), we have used the SARIMA algorithm, which works with this type of data and considers the possibility of the existence of a seasonality. Due to this last aspect, the dataset used to obtain this extra time range has been shortened to eliminate the annual periodicity and maintain the daily one.

Bearing this in mind, this algorithm is applied with the most optimal configuration of its parameters and the outcome is shown in Fig. 15. In addition, Table 6 shows the metrics used for the numerical evaluation of the method. It can be seen that a precision of 96% is obtained, which can be considered a good performance. Lastly, this final dataset reached is the one used by the ML models chosen to perform and handle the data flow environment.

**Table 6**
Metrics for the evaluation of future values estimation method.

| Technique | Mean Absolute Percentage Error (MAPE) | Mean Absolute Error (MAE) |
|---|---|---|
| SARIMA (0,1,1)(2,1,0) [24] | 0.040075 | 0.87221 |

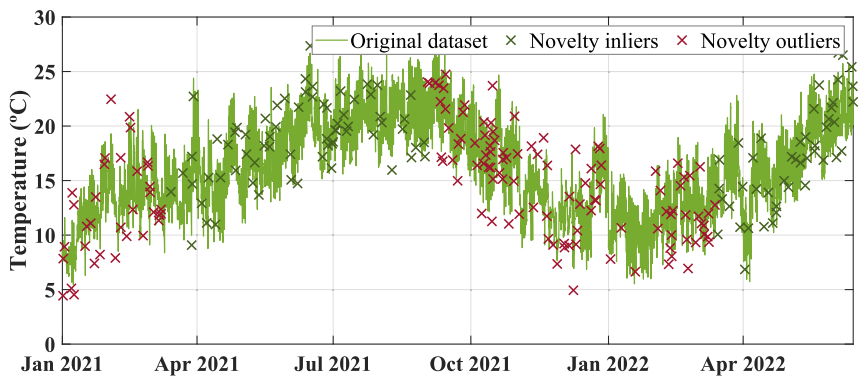

**Fig. 14.** Novelty detection on the Santander temperature dataset after anomaly cleaning and estimation of the remaining gaps with the One-Class SVM algorithm.

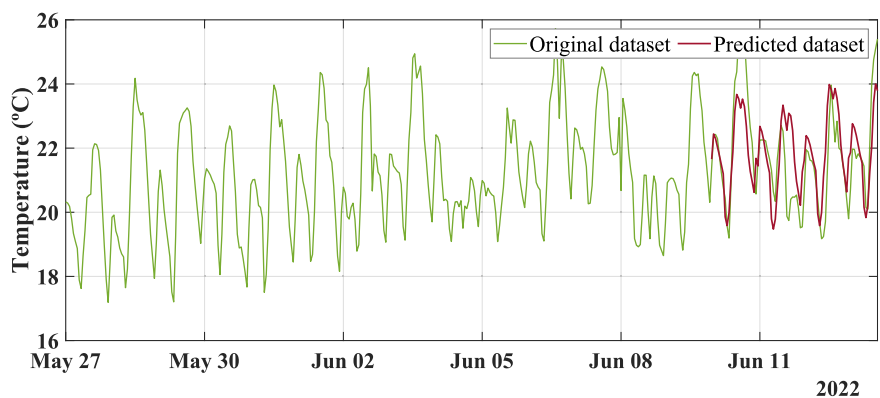

**Fig. 15.** Forecasting the immediate future with the SARIMA algorithm (configuration (0,1,1)(2,1,0) [24]).

## 6. Conclusions

With the ever-increasing importance that data has nowadays, guaranteeing its quality and being able to understand all the dimensions that DQ has, becomes an unavoidable condition for any data platform. In particular, in the case of IoT platforms, due to the particularities of IoT infrastructures, which makes sometimes impossible to impose minimum DQ requirements, this is particularly important.

In this paper we have presented the work carried out to evaluate specific DQ dimensions for IoT data streams and compensate, through AI-enabled data curation mechanisms, poor values in these dimensions. The technological contributions of this work address the gaps in existing research through:

- The mathematical formulation of the four key DQ dimensions, namely accuracy, precision, completeness and timeliness, beyond the existing theoretical definitions.
- The definition of a fifth DQ dimension on the informative axis of DQ, that provides tangible modeling for subjective DQ dimensions focused on making the data more usable.
- The design and development of mechanisms for data curation that leverages AI to improve DQ from real-time IoT data streams on a continuous basis.
- The integration of these DQ assessment and data curation mechanisms within an IoT platform leveraging linked data principles and NGSI-LD standard to maximize all the selected DQ dimensions (i.e. accuracy, precision, completeness, timeliness and usability) on the IoT data streams that such platform exposes.

Moreover, the paper presents the main results from the thorough evaluation that has been made to the implemented AI-based data curation mechanisms. Both the outlier and novelty detection mechanisms implemented as well as the data forecasting algorithm integrated within the data curation module has demonstrated adequate behavior in terms of removal of IoT observations that would reduce DQ and in terms of reduced error for synthetically generated IoT observations. The combined effect of these two mechanisms applied to a real-world IoT infrastructure like the one used for the evaluation (i.e. smart city IoT deployment in the city of Santander), results on enriched data with improved DQ available to the data consumers.

The main limitations from the proposed solution relates with the number of DQ dimensions that have been assessed and its applicability to any data stream. In this sense, it would be desirable to dynamically allow the inclusion of the metadata corresponding to each DQ dimension (with a more extensive catalogue besides the four ones described in this paper) on a per consumer basis. This way, each consumer would be able to tailor the data received instead of having it pre-established by the provider. Additionally, during the evaluation of the novelty and outlier detection mechanisms that we have implemented, we have realized how specific repetitions of outliers within a data stream might be able to make the AI-based solutions proposed to mistakenly classify outliers as non-outliers. This characteristic could be used to intentionally poison the system so that it misbehaves (just as fake news does, repeating an untrue piece of news as many times as necessary until it becomes true for a significant majority of readers).

The next steps in our research are addressing these limitations by enlarging the number of DQ dimensions assessed and enabling in the DET the capacity to dynamically add the DQ-related metadata requested by the consumer, as well as further investigating the potential weakness of the outlier detection mechanisms and proposing alternatives to make them more robust against such kind of "attacks".

## Declaration of competing interest

The authors declare that they have no known competing financial interests or personal relationships that could have appeared to influence the work reported in this paper.

## Data availability

The authors are unable or have chosen not to specify which data has been used

## Acknowledgments

This work was supported by the European Commission CEF Programme by means of the project SALTED "Situation-Aware Linked heTerogeneous Enriched Data" under the Action Number 2020-EU-IA-0274 and by the Spanish State Research Agency (AEI) by means of the project SITED "Semantically-enabled Interoperable Trustworthy Enriched Data-spaces" under Grant Agreement No. PID2021-125725OB-I00.

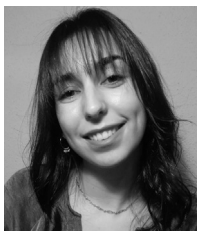

**Laura Martín** is a research fellow in the Network Planning and Mobile Communications Laboratory at Universidad de Cantabria, Spain. She received his Master's degree in Telecommunications Engineering from Universidad de Cantabria in 2022, where she is currently pursuing the Ph.D. degree in the same field. Her research interests are on the application of Artificial Intelligence for the enrichment of IoT Data. Moreover, she is also active on applying semantic web principles to data sharing and, this way, developing a fully distributed enriched data sharing ecosystem.

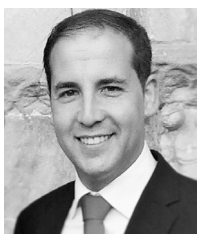

**Dr. Luis Sánchez** is Associate Professor at Universidad de Cantabria (Spain). He received M.Sc. (2002) the Ph.D. (2009) in Telecommunications Engineering. He is active on the IoT-enabled smart cities, and the application of AI for data enrichment. He has led and/or participated in more than 15 projects belonging to different EU Framework Programs. He has authored more than 60 papers at international journals and conferences. He often participates in panels discussing about innovation supported by IoT in Smart cities. He also acts as expert for several European countries national funding agencies.

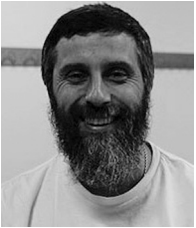

**Dr. Jorge Lanza** is an Associate Professor at the Network Planning and Mobile Communications Laboratory at the University of Cantabria (UC), Spain. He received his PhD in telecommunications engineering from University of Cantabria in 2014. He has participated in several research projects, national and international, with both private and public funding. Currently his research is focused on IoT infrastructures towards federating deployments in different locations using semantics technologies. In addition, his work has included combined mobility and security for the wireless Internet.

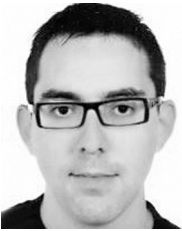

**Dr. Pablo Sotres** is a senior research fellow in the Network Planning and Mobile Communications Laboratory, which belongs to the Communications Engineering department at the University of Cantabria, Spain. He received Telecommunications Engineering degree and PhD from the University of Cantabria in 2008 and 2021 respectively. He has been involved in several different international projects framed under the smart city paradigm, such as SmartSantander; and related to inter-testbed federation, such as Fed4FIRE, Fed4FIRE+ and Wise-IoT.